\documentclass[5p,twocolumn]{myelsarticle}
\usepackage[english]{babel}
\usepackage[utf8x]{inputenc}
\usepackage{subfigure}
\usepackage{booktabs}
\usepackage{amsmath}
\usepackage{graphicx}
\usepackage[colorinlistoftodos]{todonotes}
\usepackage{algorithm2e}
\usepackage{amssymb}
\usepackage{algpseudocode}
\usepackage{calc}
\newlength{\arrowwidth}
\usepackage{microtype}
\usepackage{multirow}
\usepackage{siunitx}
\usepackage{tabularx}
\usepackage{placeins}
\usepackage{array}
\usepackage{makecell}
\usepackage{afterpage}
\usepackage[pscoord]{eso-pic}
\newcommand{\placetextbox}[3]{
  \setbox0=\hbox{#3}
  \AddToShipoutPictureFG*{
    \put(\LenToUnit{#1\paperwidth},\LenToUnit{#2\paperheight}){\vtop{{\null}\makebox[0pt][c]{#3}}}%
  }%
}%

\makeatletter
\newcommand*{\indep}{
  \mathbin{
    \mathpalette{\@indep}{}
  }
}
\newcommand*{\nindep}{
  \mathbin{
    \mathpalette{\@indep}{\not}
  }
}
\newcommand*{\@indep}[2]{
  \sbox0{$#1\perp\m@th$}
  \sbox2{$#1=$}
  \sbox4{$#1\vcenter{}$}
  \rlap{\copy0}
  \dimen@=\dimexpr\ht2-\ht4-.2pt\relax
  \kern\dimen@
  {#2}
  \kern\dimen@
  \copy0
}

\AtBeginDocument{}

\usepackage{hyperref}

\journal{Journal of Biomedical Informatics}

\makeatletter
\newcommand{\customlabel}[2]{%
   \protected@write \@auxout {}{\string \newlabel {#1}{{#2}{\thepage}{#2}{#1}{}} }%
   \hypertarget{#1}{}
}
\makeatother
\begin{document}

\begin{frontmatter}

\title{OpBerg: Discovering causal sentences using optimal alignments}
%
\author[csaddress]{Justin Wood}
%
\author[miiaddress]{Nicholas J.\ Matiasz}
%
\author[silvaadress]{Alcino J.\ Silva}
%
\author[miiaddress]{William Hsu}
%
\author[mayoaddress]{Alexej Abyzov}
%
\author[csaddress]{Wei Wang\corref{mycorrespondingauthor}}
\ead{weiwang@cs.ucla.edu}
%
\cortext[mycorrespondingauthor]{Corresponding author}
%
\address[csaddress]{Department of Computer Science, UCLA, 3551 Boelter Hall, 580 Portola Plaza, Los Angeles, CA 90095}

\address[miiaddress]{Department of Bioengineering, UCLA, 924 Westwood Blvd., Ste.\ 420, Los Angeles, CA 90024}
\address[silvaadress]{Department of Neuroscience, UCLA, 695 Charles Young Dr.\ S., Rm.\ 2357, Los Angeles, CA 90095}
\address[mayoaddress]{Center for Individualized Medicine, Department of Health Sciences Research, Mayo Clinic, 200 First St. SW Rochester, MN 55905}
\begin{abstract}
\textit{Background: } The biological literature is rich with sentences that
describe causal relations.
Methods that automatically extract such sentences can help biologists to synthesize
the literature and even discover latent relations that had not been articulated
explicitly.
Current methods for extracting causal
sentences are based on either machine learning or a 
predefined database of causal terms.
Machine learning
approaches require a large set of
labeled training data and can be susceptible to noise.
Methods based on predefined databases 
are limited by the quality of their curation and
are unable to capture new concepts or mistakes
in the input. \\
\textit{Objectives: } This paper presents
a novel and outperforming method for extracting causal relations from
text by aligning the part-of-speech (POS) representations of an input
set with that of known causal sentences. \\
\textit{Methods: } This method extracts causal
relations by
adapting and improving a method designed for a
seemingly unrelated problem:
finding alignments between genomic sequences.
Each sentence for training and testing is
converted to a representation where each word
is replaced by its corresponding POS token.
Given
a set of POS tokens labeled as causal
and non-causal,
we take an unlabeled token sequence to be of the
same class as its best aligning labeled match.
Paramount to this approach is finding the
best number of
alignments (breakpoints) along with the best 
alignment for each breakpoint.\\
\textit{Results: } The execution time of
OpBerg is compared against the state-of-the art machine learning
algorithms for the task of causality extraction using a training
set size of 100 sentences and a test size ranging from
1,000 to 10,000 sentences. OpBerg is shown
to run faster by a factor of 10 over the
compared methods. Next OpBerg is compared
against the same methods in a causality retrieval task.
The task is to correctly retrieve the causal 
statements
from a set of research articles.
Again, OpBerg significantly outperforms the competing
methods.\\
\textit{Conclusion: } Our experiments show that when applied to the task of  
finding causal sentences in biological literature,
our method improves on the accuracy of other methods
in a computationally efficient manner.
\end{abstract}


\end{frontmatter}

\section{Introduction}
Researchers who perform biological experiments convey their discovery in published research articles, which contain descriptions of causal relations.
This growing literature provides an enormous amount of information and represents the current state of biological understanding. 
This documentation of scientific discovery can 
verify previous experiments, provide insights to researchers~\cite{silva2015need}, and motivate future research~\cite{explan}.\par
\indent
These corpora of biological text
are growing at an exponential rate. Algorithms and approaches
are thus needed to extract the relevant information, allowing biologists
to understand and connect biological
processes. Since researchers describe causal
connections among biological entities in 
free-text research papers, it is logical 
to extract these connections using natural language processing (NLP).\par
\indent
A causal assertion 
can be thought of as a relation between an agent and a target.
Often in biological studies, an agent is either passively observed or actively manipulated, and
a change or lack thereof is noted in
a target. Although this type of result can
be described across many different and sometimes
nonadjacent sentences, this paper focuses only
on causal assertions appearing in a single
sentence. This approach has the advantage of limiting the
search range for descriptions of causality and
takes advantage of existing methods that can 
reliably fragment documents into collections
of sentences~\cite{stanford:nlp}.\par
\indent
Existing methods for causality
extraction use either predefined knowledge bases,
word lists, other types of databases~\cite{nonml1,nonml2,kaplan1991knowledge,DBLP:conf/flairs/GirjuM02,DBLP:journals/bmcbi/BuiNBS10}, 
or are based on statistical techniques---often some form
of machine learning~\cite{ml1,ml2,ml3,ml4,exctraction:2,extraction:1}.
Predefined knowledge bases are of course limited by the quality
of the knowledge base itself. Often, these sources
are manually curated and do not always contain
all possible words or phrases of interest.
Additionally, they require exact matches to be useful.
For instance, if a knowledge base 
contains causal verbs and a potential causal
sentence contains the misspelled verb ``cuases'' (instead of ``causes''), the sentence will be 
dismissed due to the misspelling. These predefined
knowledge bases are also not able to capture new
words or concepts, and they are not extensible to
other tasks such as extracting causality from text
in other languages.

One solution to these problems is to use
existing machine learning techniques. But these
approaches often require large amounts of
labeled training data, something that can be expensive
and tedious to obtain. These barriers of time and cost are expanded when the task is to discover
more fine-grained details pertaining to causality, 
such as that of finding the specific types of studies and
outcomes that lend evidence for a causal assertion.
Additionally, the vocabulary for biomedical free text can be quite
large, as it contains not only common words
but also domain-specific terms. This
large vocabulary set requires an even larger training data
for the machine-learning model to predict the necessary
components for representing causal phenomena.\par
\indent Thus, to automatically extract causal sentences, an
approach is required that does not suffer from limitations
in the size of the training data, and that can be performed efficiently. The approach presented in this paper is inspired by the analogy of the aforementioned problem to that of comparing  a set of genomic sequences in bioinformatics.\par
\indent
Though it may not be obvious, there is indeed a connection between aligning sequences in genomic data and finding causal sentences in free text. 
While each sentence may contain a unique set of words, the part-of-speech (POS) sequence of each sentence is likely to be much more common. 
Breaking each sentence into its grammatical structures can thus help to identify patterns in the way that causal relations are described. 
Thus, applying an alignment method to the grammatical structures of sentences has the potential to discover similarities that may be missed by approaches that focus only on words.
We further illustrate this with the following example of three sentences
and their corresponding POS mappings (for brevity we replace the POS label with a single character: A = pronoun, B = verb, C = determiner, E = adjective, F = noun, G = preposition):\par
    \begin{center}
    \includegraphics[width=258.5pt]{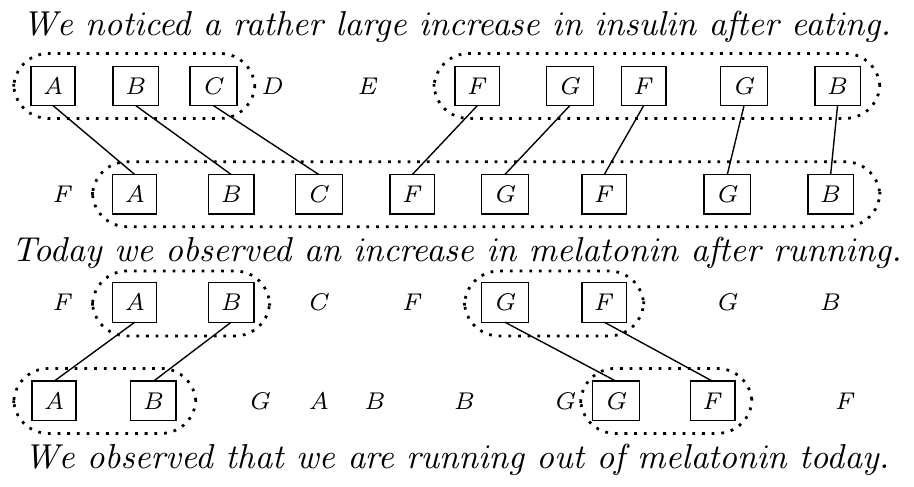}
    
\end{center}
\par
\noindent Here the first two sentences are talking about two different things; yet both
are causal sentences. Their POS structures are similar. 
In comparison, the second and third
sentence share a lot of words, more so than the first and
second sentences, yet their POS representations have fewer matching elements,
with long gaps in between matches. 
Therefore, knowing that the second sentence is causal, we
cannot determine whether the third sentence is causal. It is our hypothesis that given a labeled set of causal sentences $C+$ and non-causal sentences $C-$, a new sentence $s$ is classified as a causal sentence if its POS structure is most similar to a causal sentence (than any non-causal sentences) and the similarity ($S$) is above a   
threshold $\delta$, 
\begin{displaymath}
\max_{c \in C+} S(c, s) > \max_{c \in C-} S(c, s) \land \max_{c \in C+} S(c, s) > \delta
\end{displaymath}
\indent The approach presented here finds causal relations by comparing the POS mappings of unlabeled sentences to that of labeled sentences 
A new causal sentence is discovered by identifying the optimal number of alignments between the grammatical representations
of the sentences.
This alignment approach can thus classify causal sentences accurately and efficiently, and it has the potential to be used for other problems as well.

However, existing methods of sequence alignment are insufficient for aligning POS representations of free text: either (1) they require the user to specify the number of local alignments~\cite{age} or (2) they introduce a gap penalty for each new local alignment~\cite{gap3}, possibly leading to erroneous alignments~\cite{age}.
Given the nature of free text, it is unreasonable to ask the users to pre-specify the number of local alignments.
Here, we generalize existing alignment algorithms by removing the need to specify these parameters, while keeping the same algorithmic complexity in terms of both space and time.
This generalization allows us to efficiently apply the algorithm to NLP.
The techniques presented in this paper need not be limited to extracting
causality. We recommend using our approach for information retrieval tasks
dealing with sequential similarity when the input data set is too small to be
sufficient for machine learning.

\section{Methods}\label{app}
The proposed approach, named \textit{OpBerg}, builds upon the AGE algorithm: 
it uses a similar strategy to
find the optimal number of local alignments. AGE can
be thought of as splitting the input sequences into
segments and then running a local alignment
algorithm on those segments.
The original form of AGE that involves going forward
and reverse in two matrices makes any additional
alignment gaps difficult to compute and store. It is thus the
linear-space algorithm that
holds the key to solving the problem of optimal local alignments.
Because the directionality moves from left
to right (or right to left), this approach can be used
to split the strings into an arbitrary number of
segments. Further information is needed to
implement the proposed approaches that retain
necessary information about the locations of the gaps in the alignments. 
The change required to the original
AGE equation is the addition of a matrix
that stores the location of a newly created alignment
(for brevity we show only
the relevant addition to Equation 1):
\vspace{1mm}
\begin{displaymath}
X(i, j)=
    \begin{cases}
      X(i\text{-}1, j), & \!\text{if}\ R(i,j) = R(i\text{-}1, j) + Q \\
      X(i\text{-}1, j\text{-}1), & \!\text{if}\ R(i,j) = R(i\text{-}1, j\text{-}1) + S(a_i, b_j) \\
      X(i, j\text{-}1), & \!\text{if}\ R(i,j) = R(i, j\text{-}1) + Q \\
      (i\text{-}1, j\text{-}1), & \!\text{if}\ R(i,j) = M(i, j\text{-}1) + S(a_i, b_j) \\
      (0,0), & \! \text{if}\ R(i,j) = 0 \\
    \end{cases}
\end{displaymath}
\vspace{1mm}
\par
This optimal solution also uses
our proposed concept of score length, whose definition is as follows:\par
\textit{Definition: score length.}
The score length for the alignment of POS tokens $a_i a_{i+1}\dotso a_{i+d_1}$
and $b_j b_{j+1}\dotso b_{j+d_2}$ is defined as the difference between the
max score in the alignment matrix at cell locations $(i+d_1,j+d_2)$ and $(i,j)$.\par
\indent
A naive algorithm for solving the optimal alignment problem
is to run the existing AGE method
on every possible number of local alignments
that could reasonably occur:
\begin{displaymath}
L(i, j, 0) = Max\left\{
                \begin{array}{ll}
                  L(i-1, j, 0) + Q\\
                  L(i-1, j-1, 0) + S(a_i, b_j)\\
                  L(i, j-1, 0) + Q\\
                  0\\
                \end{array}
              \right\}
\end{displaymath}
\begin{displaymath}
L(i, j, k) = Max\left\{
                \begin{array}{ll}
                  L(i-1, j, k) + Q\\
                  L(i-1, j-1, k) + S(a_i, b_j)\\
                  L(i, j-1, k) + Q\\
                  M(i-1, j-1, k) + S(a_i, b_j)\\
                  0\\
                \end{array}
              \right\}
\end{displaymath}
\begin{displaymath}
M(i, j, 0) = Max\left\{
                \begin{array}{ll}
                  M(i-1, j, 0)\\
                  L(i, j, 0)\\
                  M(i, j-1, 0)\\
                \end{array}
              \right\}
\end{displaymath}
\begin{displaymath}
M(i, j, k) = Max\left\{
                \begin{array}{ll}
                  M(i-1, j, k)\\
                  L(i, j, k-1)\\
                  M(i, j-1, k)\\
                \end{array}
              \right\}
\end{displaymath}
\begin{displaymath}
X_I(i,j) =  X(i-1,j,k)
\end{displaymath}
\begin{displaymath}
X_M(i,j) =  X(i-1, j-1,k)
\end{displaymath}
\begin{displaymath}
X_D(i,j) =  X(i,j-1,k)
\end{displaymath}
\begin{displaymath}
X_X(i,j) =  X(i,j,k-1)\cup(i-1, j-1)
\end{displaymath}
\begin{displaymath}
X_0 =  (0, 0)
\end{displaymath}
\begin{displaymath}
L_I(i,j,k) = L(i-1, j, k) + Q
\end{displaymath}
\begin{displaymath}
L_M(i,j,k) = L(i-1, j-1,k)+S(a_i, b_j)
\end{displaymath}
\begin{displaymath}
L_D(i,j,k) = L(i, j-1,k) + Q
\end{displaymath}
\begin{displaymath}
L_X(i,j,k) = M(i-1, j-1, k)+S(a_i, b_j)
\end{displaymath}
\begin{displaymath}
X(i, j, k) =
    \begin{cases}
      X_I(i,j), & \text{if}\ L(i,j,k)=L_I(i,j,k) \\
      X_M(i,j), &\text{if}\ L(i,j,k)=L_M(i,j,k) \\
      X_D(i,j), & \text{if}\ L(i,j,k) = L_D(i,j,k) \\
      X_X(i,j), & \text{if}\ L(i,j,k)=L_X(i,j,k)\\
      X_0, & \text{if}\ L(i,j,k) = 0\phantom{space space spa}(2)\\
    \end{cases}
\end{displaymath}\par
\placetextbox{0.25}{0.75}{\customlabel{eqn:two}{2}}
Although this may
seem to be an unreasonable solution, the running
time and memory usage remain polynomial
and thus feasible for small input sizes.\par
\indent
As shown by Equation 2, the change required is to compute and store 
the possible different alignments using a separate
matrix for each split. 
A new variable is introduced, $k$, which represents the current number of
local alignments to run on the given input sequences. The
results of these additions require an $n$ factor increase
in both running time and memory retention, where $n$ is
defined as the size of the largest input POS token
sequence. The running
time becomes $\mathcal{O}(n^3)$ with memory required
as $\mathcal{O}(n^3)$.\par
\indent
Like the segmented least squares problem~\cite{least}, 
it is intuitive to add a penalty ($P$) for each additional
increase in local alignments. This penalty is needed since
otherwise, the optimal alignment would always just match
individual POS tokens. Because this penalty is
proportional to the number of local alignments, we
make the penalty a simple linear constant. The
maximum alignment score can then be defined as:
\begin{displaymath}
\underset{1 \leq k \leq n}{Max}[P \times k + M(|A|, |B|, k)]\,,
\end{displaymath}
where $A$ and $B$ are the input POS token sequences mapped
from two sentences. $M$ is the three-dimensional maximum matrix which holds the maximum
alignment score for each $a_i$, $b_j$, and $k$; where $a_i \in A$ and $b_j \in B$.\par
\par A simple linear penalty
constant reveals that returning one such alignment is
not a trivial and deterministic task. The linear penalty
can be thought of as an additional larger gap penalty, thus
taking the form of a generalized global alignment~\cite{gap3}. It has
already been shown~\cite{age} that this can lead to
improper alignments.\par
\indent
The question then becomes: What is the optimal number of alignments? For example, a user may 
prefer to find an alignment that has only
$1$ large segment aligned and a score of $28$ over $10$
alignments and a score of $29$. To determine the correct
number of alignments, this work focuses on three major
trade-offs:
\begin{enumerate}
\item Number of alignments.\vspace{-5pt}
\item Score length to break apart an alignment ($\alpha$).\vspace{-5pt}
\item Minimum score length to start an alignment ($\beta$).
\end{enumerate}\par
\indent The naive algorithm solves the problem of finding
the optimal number of local alignments, but it does so
at a considerable cost. For causal sentences, this increase
is not infeasible due to the relatively low input size
of sentences. But running this algorithm over a very
large corpus like the entirety of PubMed Central\footnote{\url{https://www.ncbi.nlm.nih.gov/pmc/}} would
carry a considerable execution cost. Thus,
it is advantageous to seek solutions that are more efficient in both time and space. Opberg, the approach we present here, seeks to reduce memory by a factor of $n^2$ and execution
time by a factor of $n^2$.
\subsection{OpBerg}
Note that during execution of the naive algorithm described above, once it is
decided that a new local alignment is a better choice, 
the optimal solution can then only be of the same or more
alignments. This allows us to reuse the existing $M$
matrix and shave off the $k$ dimension, allowing for much simpler bookkeeping. 
We introduce a new matrix $L$ that
represents the values of a local alignment.
The $M$ matrix then takes on the interpretation of
a matrix whose values are the max of the previous max
$M$ cell value and the corresponding $L$ cell value.
The optimal solution then can be in the $L$ matrix (that
is, performing a local alignment) or in the $M$ matrix (that is,
moving through the cells of the matrix and not decreasing in value).
We use the notation that if the optimal solution is in the $L$ matrix,
then it is in the ``$L$'' or ``alignment'' state; and if the optimal solution is in the $M$ matrix, then
it is in the ``$M$'' or ``max'' state.
Given that there is only
one $L$ state, it is entirely possible for the optimal solution
to transition multiple times from the $M$ state to the $L$ state before
beginning an alignment. We store the values of a transition
in a new matrix $N$ which holds the point of a transition in and out of the
$M$ state. Another matrix $X$ holds the points of all transitions through
the optimal solution.\par
\indent
The three trade-offs discussed above can be dealt with in various ways. 
To account for the number of alignments, we can leave in the
original penalty $P$, but instead of considering this as
a larger gap penalty, one can think of it as a value less
than $1$ and possibly even $0$ (with the original gap penalty
greater than $1$). By doing so, one can easily
gauge at what point a new alignment gap starts to weigh negatively
on the score and thus becomes less desirable.\par
\indent
To consider the minimum score length that is considered
to break apart an alignment, we need only consider the point
at which the algorithm exits the max state. If the current
alignment has not dropped below the input score length $\alpha$,
then we will restrict the transition until the appropriate 
threshold has been reached.\par
\indent
Likewise for the start of an alignment, with the change
only to the entering of the max state. This requires storing the
score at the start of entering the alignment state so that we
can compare the difference to see if we are above threshold. This value
is stored in the matrix $H$. This allows us to restrict the
length as we do for breaking apart an alignment, but a key
difference happens when an alternative alignment is nonexistent. 
For example, a user may prefer not to start a segment
of only $3$ matched characters unless this is the max score out of any
alternative alignments by a score of $3$ matches. We must introduce
into this restriction of a transition into the max state a way to keep track of how
a score length smaller than $\beta$ influences the score. That is,
we do not necessarily want to discard these alignments unless there
is a better alignment available. A new parameter is introduced,
$\gamma(x)$, which allows the user to specify a function to
weigh how important a certain score length is when it is below threshold, but
no higher scoring alternatives exist.\par
\indent With these parameters, the algorithm is bound to
a running time of $\mathcal{O}(n^2)$ and memory
requirements of $\mathcal{O}(n^3)$. The intuition
for this algorithm follows the intuition of segmented
least squares. In the segmented least squares problem,
we are searching for a balance between accuracy and
number of lines, whereas in OpBerg we seek this
parsimony between alignment score and number of jumps
through the matrix to start a new local alignment.
The trade-off is then enforced by the penalty constants
$P$, $\alpha$, $\beta$, and function $\gamma(x)$.
\subsubsection{Affine Gap}
It should not always be the case that insertions
and deletions (indels) between the inputs are
weighted equally, regardless of where they occur.
For instance, in certain 
causal sentences, a large cluster of indels may represent a
tangential segment of words. To capture these occurrences,
an affine gap model that takes into account segments of
tangential words
must be adapted to OpBerg.\par
\indent
The changes required of OpBerg for an affine gap are similar to those
in the original local alignment algorithm~\cite{affine}.
Three matrices---representing a match/mismatch ($L_G$), insertion ($L_I$), and deletion ($L_D$) transitions, respectively---must be used
in place of the original $L$ matrix. The max matrix $M$ cannot enter into
any of these three states because it represents a jump through the inputs, so it remains
the same. Also, since a local alignment must start and end with a match (diagonal move),
the transition between the $L$ states to the $M$ states can occur only through the
new $L_G$ matrix. This also applies to the $X$ and $N$ matrices, as they
only must monitor
jumps between the $L_G$ and the $M$ matrices.\par
\indent
The recurrent relations needed for the affine gap OpBerg model are given in their
entirety as:
\begin{displaymath}
L_I(i, j) = Max\left\{
                \begin{array}{ll}
                  L_I(i-1, j) + E \\
                  L_G(i-1, j) + O + E\\
                  L_D(i-1, j) + O + E\\
                \end{array}
              \right\}
\end{displaymath}
\begin{displaymath}
H_I(i, j) =
    \begin{cases}
      H_I(i-1, j) & \text{if}\ L_I(i, j)=L_I(i-1, j) + E \\
      H_G(i-1, j) & \text{if}\ L_I(i, j)=L_G(i-1, j) + O + E \\
      H_D(i-1, j) & \text{if}\ L_I(i, j)=L_D(i-1, j) + O + E \\
    \end{cases}
\end{displaymath}
\begin{displaymath}
\theta(i,j) =
    Max\left\{\begin{array}{ll}
      M(i-1, j)\\
      M(i, j-1)\\
    \end{array}
    \right\}
\end{displaymath}
\begin{displaymath}
\delta(i,j) = Max\left\{
                \begin{array}{ll}
                  0\\
                  L_I(i-1, j-1) + S(a_i,b_j)\\
                  L_G(i-1, j-1) + S(a_i,b_j)\\
                  L_D(i-1, j-1) + S(a_i,b_j)\\
                \end{array}
              \right\}
\end{displaymath}
\begin{displaymath}
L_{G,I,H}(i, j) = L_I(i-1, j-1) + S(a_i, b_j)
\end{displaymath}
\begin{displaymath}
L_{G,G,H}(i, j) = L_G(i-1, j-1) + S(a_i, b_j)
\end{displaymath}
\begin{displaymath}
L_{G,D,H}(i, j) = L_D(i-1, j-1) + S(a_i, b_j)
\end{displaymath}
\begin{displaymath}
L_{G,M,H}(i, j) = M(i-1, j-1) + S(a_i,b_j) + P
\end{displaymath}
\begin{displaymath}
\psi(i,j) =
    \begin{cases}
      \theta(i,j) & \text{if}\ \delta(i,j) = 0 \\
      H_I(i-1, j-1) & \text{if}\ \delta(i,j) = L_{G,I,H}(i, j)\\
      H_G(i-1, j-1) & \text{if}\ \delta(i,j) = L_{G,G,H}(i, j)\\
      H_D(i-1, j-1) & \text{if}\ \delta(i,j) = L_{G,D,H}(i, j)\\
    \end{cases}
\end{displaymath}
\begin{displaymath}
\pi(i,j) = M(i-1, j-1) + S(a_i,b_j) + P
\end{displaymath}
\begin{displaymath}
\epsilon(i,j) =
    \begin{cases}
     \pi(i,j) & \text{if}\ \delta(i,j) - \psi(i,j) \leq \alpha\\
      -\infty & \text{otherwise} \\
    \end{cases}
\end{displaymath}
\begin{displaymath}
L_G(i, j) =
    Max\left\{\begin{array}{ll}
      \delta(i,j)\\
      \epsilon(i,j)\\
    \end{array}
    \right\}
\end{displaymath}
\begin{displaymath}
H_G(i, j) =
    \begin{cases}
      \theta(i,j) & \text{if}\ L_G(i, j) = 0 \\
      H_I(i-1, j-1) & \text{if}\ L_G(i, j) = L_{G,I,H}(i, j) \\
      H_G(i-1, j-1) & \text{if}\ L_G(i, j) = L_{G,G,H}(i, j) \\
      H_D(i-1, j-1) & \text{if}\ L_G(i, j) = L_{G,D,H}(i, j) \\
      \theta(i,j) & \text{if}\ L_G(i, j) = L_{G,M}(i, j) \\
    \end{cases}
\end{displaymath}
\begin{displaymath}
L_D(i, j) = Max\left\{
                \begin{array}{ll}
                  L_I(i, j-1) + O + E\\
                  L_G(i, j-1) + O + E\\
                  L_D(i, j-1) + E\\
                \end{array}
              \right\}
\end{displaymath}
\begin{displaymath}
L_{D,I,H}(i, j) = L_I(i, j-1) + O + E
\end{displaymath}
\begin{displaymath}
L_{D,G,H}(i, j) = L_G(i, j-1) + O + E
\end{displaymath}
\begin{displaymath}
L_{D,D,H}(i, j) = L_D(i, j-1) + E
\end{displaymath}
\begin{displaymath}
H_D(i, j) =
    \begin{cases}
      H_I(i, j-1) & \text{if}\ L_D(i, j) = L_{D,I,H}(i, j)\\
      H_G(i, j-1) & \text{if}\ L_D(i, j) = L_{D,G,H}(i, j)\\
      H_D(i, j-1) & \text{if}\ L_D(i, j) = L_{D,D,H}(i, j)\\
    \end{cases}
\end{displaymath}
\begin{displaymath}
\zeta(i,j) =
    \begin{cases}
      L_G(i, j) & \text{if}\ L_G(i, j) \geq \beta\\
      \gamma(L_G(i, j)) & \text{otherwise} \\
    \end{cases}
\end{displaymath}
\begin{displaymath}
M(i, j) = Max\left\{
                \begin{array}{ll}
                  \zeta(i,j) \\
                  M(i-1, j)\\
                  M(i, j-1)\\
                \end{array}
              \right\}
\end{displaymath}
\begin{displaymath}
L_{G,I,X}(i, j) = L_I(i-1, j) + Q
\end{displaymath}
\begin{displaymath}
L_{G,G,X}(i, j) = L_G(i-1, j-1) + S(a_i,b_j)
\end{displaymath}
\begin{displaymath}
L_{G,D,X}(i, j) = L_D(i, j-1) + Q
\end{displaymath}
\begin{displaymath}
N_{X}(i, j) =  N(i-1, j-1) \cup (i, j)
\end{displaymath}
\begin{displaymath}
X_{D}(i, j) =  X(i-1, j-1)
\end{displaymath}
\begin{displaymath}
X(i, j) =
    \begin{cases}
      X(i-1, j) & \text{if}\ L_G(i, j)=L_{G,I,X}(i, j) \\
      X_{D}(i, j) & \text{if}\ L_G(i, j) = L_{G,G,X}(i, j) \\
      X(i, j-1) & \text{if}\ L_G(i, j-1) = L_{G,D,X}(i, j) \\
      N_{X}(i, j) & \text{if}\ L_G(i, j) = \epsilon(i,j) \\
      \varnothing & \text{if}\ L_G(i, j) = 0\\
    \end{cases}
\end{displaymath}
\begin{displaymath}
N(i, j) =
    \begin{cases}
      X(i, j) \cup (i, j), & \text{if}\ M(i, j) = \zeta(i,j) \\
      N(i-1, j), & \text{if}\ M(i, j) = M(i-1, j) \\
      N(i, j-1), & \text{if}\ M(i, j) = M(i, j-1)\phantom{spa}(3)\\
    \end{cases}
\end{displaymath}
\placetextbox{0.75}{0.90}{\customlabel{eqn:three}{3}}
where $(i,j)$ represents the cell location of both matrices and
the $i$th POS token in $A$ and the $j$th POS token in $B$. $S$ is a function that takes in two POS tokens and
returns a score value. The opening gap penalty is represented
by $O$ and the extension penalty by $E$.\par
\indent Even with the newly created matrices and additional processing that
must take place to populate the matrices, the running time will
be $\mathcal{O}(n^2)$, with memory as $\mathcal{O}(n^2)$.
\section{Conclusion}
This paper introduces a novel approach to causality
discovery by considering alignments among POS mappings
of sentences. This
approach considers restrictions on the score size to break
apart an alignment and enforces a minimum length requirement
while also considering the number of alignments. OpBerg
discovers meaningful alignments that return from alignment query results that are more useful in 
finding semantic similarity of two causal sentences.
 The improved model and efficient implementation make 
OpBerg the best model to use when performing tasks that
involve the alignment of two or more sets of input,
particularly in that of POS mappings
for causal extraction.
\section*{References}
\bibliography{opberg}

\begin{thebibliography}{10}
\expandafter\ifx\csname url\endcsname\relax
  \def\url#1{\texttt{#1}}\fi
\expandafter\ifx\csname urlprefix\endcsname\relax\def\urlprefix{URL }\fi
\expandafter\ifx\csname href\endcsname\relax
  \def\href#1#2{#2} \def\path#1{#1}\fi

\bibitem{silva2015need}
A.~J. Silva, K.-R. M{\"u}ller, The need for novel informatics tools for
  integrating and planning research in molecular and cellular cognition,
  Learning and Memory 22~(9) (2015) 494--498.

\bibitem{explan}
N.~J. Matiasz, et~al., Computer-aided experiment planning toward causal
  discovery in neuroscience, Frontiers in Neuroinformatics 11 (2017) 12.
\newblock \href {http://dx.doi.org/10.3389/fninf.2017.00012}
  {\path{doi:10.3389/fninf.2017.00012}}.

\bibitem{stanford:nlp}
C.~D. Manning, et~al., The {Stanford} {CoreNLP} natural language processing
  toolkit, in: Association for Computational Linguistics (ACL) System
  Demonstrations, 2014, pp. 55--60.

\bibitem{nonml1}
M.~Selfridge, \href{http://dx.doi.org/10.1080/08839518908949924}{Toward a
  natural language-based causal model acquisition system}, Applied Artificial
  Intelligence 3~(2-3) (1989) 191--212.
\newblock \href {http://dx.doi.org/10.1080/08839518908949924}
  {\path{doi:10.1080/08839518908949924}}.
\newline\urlprefix\url{http://dx.doi.org/10.1080/08839518908949924}

\bibitem{nonml2}
B.~T. Low, K.~Chan, L.~Choi, M.~Chin, S.~Lay,
  \href{https://doi.org/10.1007/3-540-45357-1\_15}{Semantic expectation-based
  causation knowledge extraction: {A} study on hong kong stock movement
  analysis}, in: Knowledge Discovery and Data Mining - {PAKDD} 2001, 5th
  Pacific-Asia Conference, Hong Kong, China, April 16-18, 2001, Proceedings,
  2001, pp. 114--123.
\newblock \href {http://dx.doi.org/10.1007/3-540-45357-1\_15}
  {\path{doi:10.1007/3-540-45357-1\_15}}.
\newline\urlprefix\url{https://doi.org/10.1007/3-540-45357-1\_15}

\bibitem{kaplan1991knowledge}
R.~M. Kaplan, G.~Berry-Rogghe, Knowledge-based acquisition of causal
  relationships in text, Knowledge Acquisition 3~(3) (1991) 317--337.

\bibitem{DBLP:conf/flairs/GirjuM02}
R.~Girju, D.~I. Moldovan, Text mining for causal relations, in: Proceedings of
  the Fifteenth International Florida Artificial Intelligence Research Society
  Conference, May 14-16, 2002, Pensacola Beach, Florida, {USA}, 2002, pp.
  360--364.

\bibitem{DBLP:journals/bmcbi/BuiNBS10}
Q.~Bui, et~al., Extracting causal relations on {HIV} drug resistance from
  literature, {BMC} Bioinformatics 11 (2010) 101.
\newblock \href {http://dx.doi.org/10.1186/1471-2105-11-101}
  {\path{doi:10.1186/1471-2105-11-101}}.

\bibitem{ml1}
P.~Tapanainen, T.~J{\"{a}}rvinen,
  \href{http://aclweb.org/anthology-new/A/A97/A97-1011.pdf}{A non-projective
  dependency parser}, in: 5th Applied Natural Language Processing Conference,
  {ANLP} 1997, Marriott Hotel, Washington, USA, March 31 - April 3, 1997, 1997,
  pp. 64--71.
\newline\urlprefix\url{http://aclweb.org/anthology-new/A/A97/A97-1011.pdf}

\bibitem{ml2}
R.~Girju, et~al., A knowledge-rich approach to identifying semantic relations
  between nominals, Inf. Process. Manage. 46~(5) (2010) 589--610.
\newblock \href {http://dx.doi.org/10.1016/j.ipm.2009.09.002}
  {\path{doi:10.1016/j.ipm.2009.09.002}}.

\bibitem{ml3}
Q.~Do, et~al., Minimally supervised event causality identification, in:
  Proceedings of the 2011 Conference on Empirical Methods in Natural Language
  Processing, {EMNLP} 2011, 27-31 July 2011, John McIntyre Conference Centre,
  Edinburgh, UK, {A} meeting of SIGDAT, a Special Interest Group of the {ACL},
  2011, pp. 294--303.

\bibitem{ml4}
F.~Huang, A.~Yates, Open-domain semantic role labeling by modeling word spans,
  in: {ACL} 2010, Proceedings of the 48th Annual Meeting of the Association for
  Computational Linguistics, July 11-16, 2010, Uppsala, Sweden, 2010, pp.
  968--978.

\bibitem{exctraction:2}
D.~Chang, K.~Choi, Causal relation extraction using cue phrase and lexical pair
  probabilities, in: Natural Language Processing - {IJCNLP} 2004, First
  International Joint Conference, Hainan Island, China, March 22-24, 2004,
  Revised Selected Papers, 2004, pp. 61--70.

\bibitem{extraction:1}
Causal Relation Extraction.

\bibitem{age}
A.~Abyzov, M.~Gerstein, Age: defining breakpoints of genomic structural
  variants at single-nucleotide resolution, through optimal alignments with gap
  excision, Bioinformatics 27~(5) (2011) 595--603.

\bibitem{gap3}
X.~Huang, K.-M. Chao, A generalized global alignment algorithm, Bioinformatics
  19~(2) (2003) 228--233.

\bibitem{least}
R.~Bellman, et~al., Some numerical experiments using newton's method for
  nonlinear parabolic and elliptic boundary-value problems, Communications of
  the ACM 4~(4) (1961) 187--191.

\bibitem{affine}
O.~Gotoh, An improved algorithm for matching biological sequences, Journal of
  molecular biology 162~(3) (1982) 705--708.

\end{thebibliography}

\end{document}